# Individual Level Culture Effects on Multi-Perspective iTrust in B2C E-commerce

**Osama Sohaib**
Faculty of Engineering and Information Technology
University of Technology Sydney
Sydney, Australia
Email: osama.sohaib@uts.edu.au

**Kyeong Kang**
Faculty of Engineering and Information Technology
University of Technology Sydney
Sydney, Australia
Email: kyeong.kang@uts.edu.au

## ABSTRACT

Consumer trust is one of the key obstacles to online vendors seeking to extend their consumers across cultures. This research identifies culture at the individual consumer level. Based on the Stimulus–Organism–Response (S–O–R) model, this study focuses on the moderating role of uncertainty avoidance culture value on privacy and security as cognition influences, joy and fear as emotional influences (Stimuli), and individualism-collectivism on social networking services as social influence and subsequently on interpersonal trust (cognitive and affect-based trust) (Organism) towards purchase intention (Response). Data were collected in Australia and the Partial least squares (PLS) approach was used to test the research model. The findings confirmed the moderating role of individual level culture on consumer's cognitive and affect-based trust in B2C e-commerce website with diverse degrees of uncertainty avoidance and individualism.

## INTRODUCTION

A website is the central way a business-to-consumer (B2C) e-commerce firm connects with their consumers. In order to bring a sense of consumer trust in e-commerce context, online stores focus on positive online purchasing experiences for potential buyers. Therefore, the assessment of website quality is a serious measure of understanding whether the online store is providing the quality of services and interaction desired by consumers (Cheng et al. 2009). Trust is one of the key obstacles to online vendors in the success of e-commerce. Trust refers to the consumer's expectation that the online vendor will execute specific activities, regardless of the consumer's ability to control the vendor action (Pavlou and Chai 2002). The transaction complexity in e-commerce environment makes condition more uncertain, then the need for interpersonal trust grows, interpersonal trust refers to the individual trust formed in another specific party (McKnight and Chervany 2001). In e-commerce environment, the two participating parties are the online buyer and the online vendor (Tan and Sutherland 2004). The literature typically differentiates two broad foundations of interpersonal trust as cognitive and affect-based trust, cognitive-based trust is built on the available knowledge and good reasons for decision making, whereas affect-based trust is built on the emotional ties between parties (Karimov et al. 2011; McAllister 1995; McKnight et al. 1998). In addition, researchers have also found that the buyers trust formation varies across cultures (Cyr 2013; Li et al. 2011; Sia et al. 2009). However, country has been used as a proxy for culture at a group level such as by (Cyr 2008; Kim 2005; Lee et al. 2007). While online purchasing is not only an individual oriented, but in fact business-to-consumer (B2C) refers to the e-commerce type in which business sells directly to individual shoppers (Bidgoli 2002). McCoy et al. (2005) explored that considering nations as a culture is inappropriate to use because members of a group needs not to have the same cultural values. Researchers such as (Srite and Karahanna 2006; Yoon 2009) have empirically established moderator analysis with the individuals' cultural values towards online trust and purchasing intentions. Srite and Karahanna (2006) discussed that national culture is a macro-level phenomenon whereas online purchasing is a one person action; therefore measuring culture at the individual level is most appropriate which in turn effect technology acceptance. Therefore, culture analysis at the individual level is most appropriate for building consumer trust in B2C e-commerce.

This research addresses the shortcomings in the existing literature, by evaluating how cognitive, social and emotional responses to consumer interpersonal trust in B2C online store differ at the individual consumer level in Australia. We applied Individualism (IDV) and uncertainty avoidance (UA) cultural values (Hofstede 1980) at the individual level because it is believed to be highly relevant to trust in cross-culture business relationships (Cyr 2013; McCoy et al. 2005).

We adopted the Stimulus–Organism–Response (S–O–R) approach to understand the formation of cognitive and affective influences of online buyers towards purchasing intention in a B2C context. The S–O–R model suggests that environmental cues (Stimuli) effect an individual's cognitive and affective reaction (Organism) that further influences individual's behavior (Response) (Mehrabian and Russell 1974). Using S-O-R framework, the same stimuli can create different responses with differing cultural values. Therefore, this study focuses on the moderating role of IDV on social networking services, UA on privacy awareness and security as cognition influences, joy and fear as emotional influences (Stimuli) and subsequently on interpersonal trust (cognitive and affect-based trust) (Organism) towards purchase intention (Response). Following the above-mentioned, our research question is: Does the national culture value of Uncertainty Avoidance (UA) and Individualism (IDV) influences consumer's cognition, emotion, social and online interpersonal trust (iTrust) at the individual level towards purchasing intention in Australian B2C e-commerce?

## BACKGROUND and RELATED STUDIES

### Culture Differences in E-commerce

Hofstede et al. (2010) defines "culture as mental software" that is "the collective programming of the mind which distinguishes the members of one group or category of people from another". Hofstede's (1980) cultural dimensions show Australia on individualism-collectivism (IDV) with an index score (IDV: 90), uncertainty avoidance (UA) (UA: 51), which means a high individualistic and fairly uncertainty avoidance country. Individualist societies focus on individual decision making while collectivist societies focuses on group norms. Uncertainty avoidance refers to the societies, which have certain degree of uncertainty situation, and tries to avoid them. Considering culture in e-commerce is important for businesses to target consumers across the globe (Kang 2010). Hofstede's (1980/1991) cultural aspects have been comprehensively studied in B2C e-commerce research studies (An and Kim 2008; Ganguly et al. 2010; Kim 2005; Sinkovics et al. 2007; Teo and Liu 2007; Yoon 2009). Sohaib and Kang (2014) investigated cultural aspects of Australian B2C websites and the finding showed that there are differences in the representation of functional and hedonic aspects on the B2C websites across cultures (Sohaib and Kang 2015).

### iTrust in E-Commerce

Interpersonal trust (iTrust) denotes the individual trust made in another party (McKnight and Chervany 2001). In B2C e-commerce environment, the two parties are the online buyer and the online vendor (Tan and Sutherland 2004). Two aspects of interpersonal trust are cognitive and affect-based aspects (Johnson and Grayson, 2005). Cognitive-based trust develops from a "pattern of careful rational thinking and thus it reflects the customer's confidence that an e-retailer is honest, accurate, and dependable and keeps promises", whereas affect-based trust also called emotional trust, "develops from one's instincts, intuition, or feelings concerning whether an individual, group or organization is trustworthy" (Brengman and Karimov 2012). Previous research have studied cognitive-based trust determinants (such as, perceived security and privacy awareness, system reliability, information quality and coherence etc.) and affect-based trust aspects (such as mystery, joy and fear, presence of third party seal, reputation, word-of- mouth, referral, variety etc.) in e-commerce (Eastlick and Lotz 2011; Johnson and Grayson 2005; Kim 2005; Kim et al. 2008; Lee and Kozar 2010; Li et al. 2011). The findings showed that these factors have significant effects towards purchasing intention. Cognitive-based trust is a buyer confidence to rely on website design along with social networking services (Chelule 2010), while affect-based trust is the satisfying experience (feelings) demonstrated by the website itself (Éthier et al. 2008).

### Security, Privacy and Emotions in E-Commerce

Consumers are usually concerned about their privacy and security when engaged in online shopping (Salo and Karjaluoto 2007). Consumers are willing to disclose their personal information to online vendors when reliability and credibility are recognized; and hence reduces consumers' worries of privacy and security and helps build online trust toward the websites (Chen and Barnes 2007). Greenberg et al. (2008) explored that security/privacy is interrelated with website design such that when consumers are new to a website; their judgment of security/privacy is based on design features such as the professional look and feel of the website. For-example, a lack of perceived security is a major concern of many potential consumers to shop online because of general perception of risks involved in transmitting sensitive information, such as credit card payments and privacy shared information (Éthier et al. 2006; Greenberg et al. 2008). Moreover, a website appearance also encourages or discourages a consumer's online purchasing intention (Cyr et al. 2005), which in turn their emotional responses such as joy and fear (Li et al. 2011). According to (Li et al. 2011) Joy is "An emotional state of pleasure" and Fear is "An emotional state of anxiety". Current research suggests that emotions follow cognition because they are developed as a result of consumers' assessment of the risk associated with online shopping such as privacy and perceived security (Cheng et al. 2009). A number of research studies have found that consumer emotions such as joy and fear play a significant role in consumer purchasing behaviour, such as in trust beliefs and purchase intentions (Cheng et al. 2009; Salo and Karjaluoto 2007; Sheng and Joginapelly 2012). Shaver et al. (1987) identified five basic types of emotions: love, joy, anger, sadness and fear. Love, anger and sadness makes less suitable for emotional reactions to a website. Considering the B2C context, we omitted love, anger and sadness.

### Social Networking Services in E-commerce

Social networking services denotes set of actors (people/organizations) and the set of connections among the actors representing some relationship (friendship/affiliation/information exchange) (Grabner-Kräuter 2009). The web now enables collaboration through social networking services (such as online discussion forums, blogs, social networks sites etc.) among consumers. Such applications are also significantly changing the relationship between consumers and e-retailers (Lee et al. 2011). For-example, by creating its official page on social networks sites like Facebook, Twitter; the company can provide an opportunity to get closer to their offers through interactions between consumers (Thabet and Zghal 2013). In Australia, according to 2012 digital media research, "29% of all online shoppers in Australia currently follow an online shopping site on Facebook, and 21% has received a message or recommendation from a friend in the prior year on Facebook, whilst 14% has made a recommendation". More than seventy- two per cent of Internet users are engaged in a social media website, so there is a need to create a social media presence for e-commerce website (Najjar 2011). According to (Brengman and Karimov 2012) the "presence of social network application in a web interface may allow users to experience others as being psychologically present in the virtual environment. Gefen and Straub (2004) presented the influence of social presence on consumer trust such as ability, integrity, predictability and benevolence in B2C e-commerce.

## RESEARCH MODEL

Existing research shows that the biggest barrier to successful e-commerce is online consumer trust. A variety of trust models have been identified in the literature in the context of e-commerce, with each model usually focusing on a specific type of trust. However, there are similarities between trust models, such as consumer cognitive and affective reactions towards e-commerce website and purchasing online. These similarities can be grouped together to form a trust model that brings together the significant components of trust in a B2C relationship. However, even though there are similarities in the online trust models concerning the relationship between consumers and Internet stores, the existing models of online trust ignore the role of interpersonal trust.

This research aim to develop a multi-perspective trust model to identify a base set of constructs, which can be used to form a consumer interpersonal trust that can be applied to B2C e-commerce. The fast development of Internet technologies and trends of continuous growth in e-commerce creates opportunities for businesses. With e-commerce, the development of interpersonal trust between consumers and Internet stores is now possible. For example, Web 2.0 encourages consumer to interact online, and the evolving Internet provides an opportunity for existing online trust models to be updated in the B2C e-commerce context. Therefore this study aims to contribute to existing knowledge by proposing an updated multi- perspective trust model that includes the role of interpersonal trust (iTrust).

We used the Stimulus–Organism–Response (S–O–R) model to identify both cognitions and emotions-based factors in order to determine the cognitive and affect-based trust towards purchasing intention. Stimulus-organism-response (S-O-R) paradigm was first proposed by (Mehrabian and Russell 1974) in the context of environmental psychology. The paradigm suggests that stimuli from environments influence an individual's cognitive and affective reactions, which in turn lead to some response and behaviour. This paradigm was later extended and has been extensively applied to shopping outcomes of online stores. In the context of online retailing, the stimulus (S) is defined as "the total sum of all the cues that are visible and audible to the online shopper". These internal states (Organism) affect buyer responses, such as purchase intention (Response). Cognitive and affective signs is present in B2C website (Karimov et al. 2011). The individualism-collectivism (IDV) and uncertainty avoidance (UA) cultural values are adopted in this study because online shopping is an individual decision making process that comprises inherent uncertainty (Karahanna et al., 2013). Therefore, IDV and UA are closely associated with consumer trust in e-commerce. Figure 1 shows our research model.

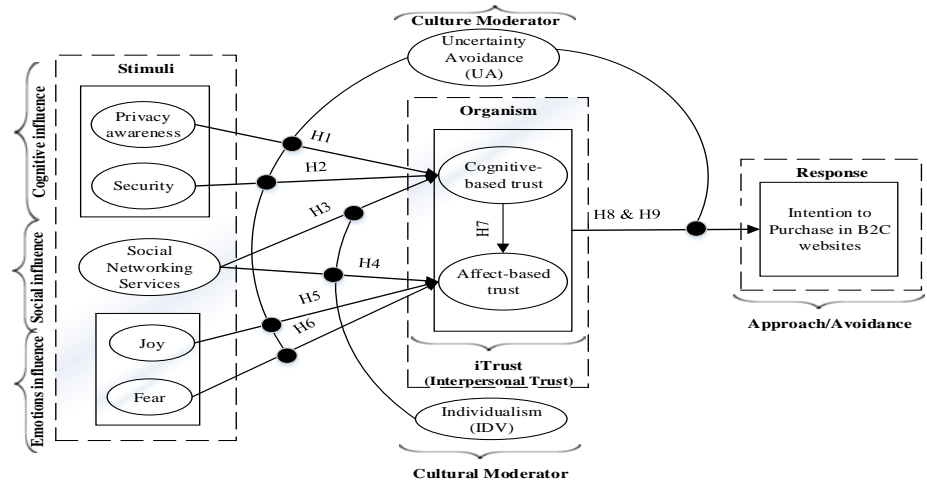

*Figure 1: Research model*

## Hypothesis Development

Privacy policies in online store often remain invisible to buyers, who rarely make the effort to read and understand those information (Tsai et al. 2011). Although a more noticeable display of privacy policy information will help buyers in their online purchasing decisions. E-vendors are responsible for protecting the information by implementing privacy policies. Concerns about privacy are likely to decrease consumer trust in an e-vendor and lower purchase intentions (Kim 2005). According to (Hofstede 2013) in high UA cultures "security is an important element in individual motivation". Cognitive-based trust determinants also vary across cultures. For-example, trust factors such as security and privacy concerns in e-commerce vary from a high uncertainty avoidance to low uncertainty avoidance type of culture (Kim and Kim 2004). (Kim 2005) noted that privacy is more significant in an individualistic-low uncertainty avoidance culture than a collectivist-high uncertainty avoidance society. Greenberg et al. (2008) also examined that privacy is considered a higher risk of violation in individualistic-low uncertainty avoidance culture while security is considered a higher risk of violation in collectivist-high uncertainty avoidance culture. Therefore,

*H1: Individual level uncertainty avoidance will moderate the relationship between privacy and cognitive-based trust such that the relationship is stronger for online consumers in Australia.*

*H2: Individual level uncertainty avoidance will moderate the relationship between security and cognitive-based trust such that the relationship is stronger for online consumers in Australia.*

According to (Swamynathan et al. 2008), users who engage in social networking, those who transact with friends of friends generally obtain significantly benefits in the form of higher user satisfaction. With the global popularity of social networking services, B2C e-commerce websites are integrating social networking elements to reach their target customers and achieve their business goals effectively across cultures (Sun 2011). Social influence can build buyer trust in online shopping in the individualistic culture (Lee et al. 2011). For-example, people in individualist cultures such as Australians are more likely to take independent decisions. (Srite and Karahanna 2006) discussed individuals with individualistic cultural values are less concerned about the views of others in their social setting. Contrary, individuals with collectivistic cultural values will conform to the views of others in a group. Thus, consumers who are individualistic avoid getting into groups and perform the transaction on their own, without requesting anybody for any help. Therefore,

*H3: Individual level IDV will moderate the relationship between social networking services and cognitive-based such that the relationship is stronger for consumers in Australia.*

*H4: Individual level IDV will moderate the relationship between social networking services and affect-based trust such that the relationship is stronger for consumers in Australia.*

In the psychology and IS literature, emotions have been empirically found to influence trust. For instance, "happiness and gratitude – emotions with positive valence – increase trust. Anger – an emotion with negative valence – decreases trust" (Dunn and Schweitzer 2005). In the B2C context, buyers may bring different emotions, and that into online shopping on the web itself may induce such emotions. For example, a visually appealing website may trigger joy, while a poorly designed website may trigger frustration or fear. Emotions have been empirically found to influence trust such that positive emotion (joy) increases trust while a negative emotion (fear) decreases trust (Li et al. 2011; Parboteeah et al. 2009). Buyers' affective reactions are taken as their emotional responses (e.g. joy and fear) to a website overall look and feel. According to (David and Hyi 2012) culture also influence the relative intensity of emotional experiences. Concerning the online shopping context, (Kim 2005) highlighted that affect-based trust (emotional-based trust) determinants are more important in collectivist cultures because people in the collectivistic society need more emotional cues to build trust relationship.

*H5: Individual level uncertainty avoidance will moderate the relationship between the relationship between initial joy and affect-based trust such that the relationship is stronger for consumers in Australia.*

*H6: Individual level uncertainty avoidance will moderate the relationship between initial fear and affect-based trust such that the relationship is stronger for consumers in Australia.*

Cognitive barriers are more serious than other kinds of barriers, especially in developing countries (Kshetri 2007). The cognition cues can also be used to induce affect-based trust towards the website (Karimov et al. 2011). Affect-based trust is "based on affect experienced from interacting with the service provider" and based on emotions (Johnson and Grayson 2005; Karimov et al. 2011; Lee et al. 2010). Cognitive-based trust is a buyer confidence to rely on website design (Chelule 2010), while affect-based trust is the satisfying experience revealed by the website (Éthier et al. 2008). Cognitive-based trust should exist before the affect-based trust develops (Johnson and Grayson 2005), and online buyers do not use them independently (Hansen 2005). Concerning the B2C e-commerce, the website interface features act as factors of the cognitive appraisal that trigger emotions (Éthier et al. 2008). (Cheng et al. 2009) also indicated that cognition has significant effects over consumers' emotions.

*H7: Cognitive-based trust has a stronger influence on affect-based trust for consumers in Australian B2C e-commerce.*

Online shopping certainly includes more uncertainties than in brick-and-mortar commerce. Moreover, acceptance rate of online purchasing is higher for individualist societies than for collectivist (Lim et al. 2004). (Yoon 2009) also noted those individualistic cultures are typically more willing to purchase online than a collectivist culture. Whereas, consumer's in a high UA culture have less influence towards intention to purchase. Therefore, it is reasonable to propose that consumers in high UA societies are more anticipated to resist shopping online than consumers in low UA society.

*H8: Individual level uncertainty avoidance will moderate the relationship between cognitive-based trust and purchase intention such that the relationship is stronger for consumers in Australia.*

*H9: Individual level uncertainty avoidance will moderate the relationship between affect-based trust and purchase intention such that the relationship is stronger for consumers in Australia.*

## APPROACH

For the validation and testing the hypotheses, data were collected from university students in Australia. University students are composed of the majority of online users and the Internet usage is comparatively higher than other aged group. Therefore, adopting students as sample is considered more applicable to online consumers (Chen and Barnes 2007). Participants were recruited from University of Technology Sydney. To ensure participant culture it is important to determine that each participant lived the most of their lives in the country and spoke the native language as their primary language (Cyr 2013). This study used an online survey methodology where participants were asked to visit a local unfamiliar online vendor. A well-localized retailer website are considered appropriate to the culture and most workable option for research (Cyr 2013). The participants were asked to visit a B2C website (oo.com.au). The participants were instructed to assume that they are interested in buying a product (1) search for any product and gather certain details to become more familiar with the chosen website (Chen and Barnes 2007) and (2) then to go through the entire online buying process up to, but excluding the clicking of the buy button to purchase the product. After the interaction with the chosen website, respondents were required to fill-out a closed-ended questionnaire on seven - point Likert scale. This includes options such as (1) strongly disagree to (7) strongly agree. Multi-scale items using at least three observable indicators measured all constructs. Previously validated survey instruments were revised and used in order to ensure the measures are adequate and representative. Furthermore, an expert from Faculty of Engineering & IT, University of Technology Sydney was used to validate whether the complete survey instrument adequately measures each construct.

Security and privacy measures derived from (Casaló et al. 2011; Kim 2005), joy and fear (Li et al. 2011; Parboteeah et al. 2009), cognitive and affect-based trust measures derived from (Karimov et al. 2011; Kim 2008) and purchase intention measures obtained from (Chen and Barnes 2007; Yoon 2009), and social networking services derived from (Hasslinger et al., 2007, Huynh and Andrade, 2012, Brengman and Karimov, 2012). Concerning the cultural instrument, Hofstede (2001) advised that his cultural instrument couldn't be used to test relationship at the individual level because the measurement items address issues from people's point of view working for IBM organization not from individual's point of view. Fortunately, (Dorfman and Howell 1988) provided an encouraging cultural instrument at the individual level based on Hofsetde cultural dimensions. The researchers (Hwang and Lee 2012b; McCoy et al. 2005; Srite and Karahanna 2006; Yoon 2009) found the Hofstede culture scales at the individual level acceptable in information systems research (more specifically in e-commerce research). Therefore, the individualism and uncertainty avoidance scales were taken and thereby adding to the items' validity.

## DATA ANALYSIS

Data collection lasted from September 2014 to December 2014. A total of 270 responses from Australia were collected. After removing incomplete responses, a total of 255 samples were used to test the proposed model. 55% respondent were males and 45% was females, 87% respondents had Internet experience 7 years or above, 58% respondents have online purchasing experience between 1-3 years.

The Partial least squares (PLS) approach (using smartPLS version 3.0) was used to test the research model. PLS is considered appropriate as it allows researchers to simultaneously assess measurement model parameters and structural path coefficients. Furthermore, it allows both formative and reflective constructs to be tested together (Chin et al. 2003). In our research model, security, privacy, emotions, cognitive-based trust, affect-based trust and purchase intention were modelled as reflective indicators because they were viewed as effects of latent variables (Hwang and Lee 2012a; Kim 2005). Whereas social networking services is a formative in nature because it is a multidimensional construct, which cover various referent groups such as social networks, online help, reviews and rankings.

### Reliability and Validity Assessment

The measurement models in PLS were evaluated by examining internal consistency, convergent validity and discriminant validity. Table 1 shows the Cronbach's reliability, composite reliability and the AVE of all constructs values exceeds the recommended value of 0.70. A social networking service is a formative construct that cannot be analyzed in this procedure. For formative indicator (social networking services), the validity of construct using outer weights was significant (p value < 0.05). In addition to this, to determine the reliability for formative indicators, the variance inflation factor (VIF) value was less than 5, which means there is no multicollinearity. An independent t-test was conducted to compare the differences concerning culture values (IDV and UA), which was significant at p<0.05.

|  | AVE | Calpha | CR | ATrust | CTrust | FEAR | IDV | INT | JOY | PRV | SNS | SEC | UA |
|---|---|---|---|---|---|---|---|---|---|---|---|---|---|
| **ATrust** | 0.81 | 0.89 | 0.93 | **0.91** | | | | | | | | | |
| **CTrust** | 0.76 | 0.78 | 0.87 | 0.26 | **0.84** | | | | | | | | |
| **FEAR** | 0.80 | 0.85 | 0.86 | -0.19 | -0.04 | **0.95** | | | | | | | |
| **IDV** | 0.88 | 0.88 | 0.88 | 0.60 | -0.00 | -0.23 | **0.97** | | | | | | |
| **INT** | 0.87 | 0.89 | 0.84 | 0.62 | 0.07 | -0.23 | 0.60 | **0.92** | | | | | |
| **JOY** | 0.73 | 0.81 | 0.87 | -0.04 | -0.04 | 0.01 | -0.02 | -0.06 | **0.85** | | | | |
| **PRV** | 0.84 | 0.86 | 0.85 | 0.51 | -0.08 | -0.27 | 0.87 | 0.59 | -0.02 | **0.91** | | | |
| **SNS** | NA | NA | NA | -0.12 | 0.09 | -0.11 | -0.00 | -0.06 | -0.07 | 0.00 | **1.00** | | |
| **SEC** | 0.71 | 0.85 | 0.89 | -0.02 | -0.05 | 0.073 | -0.07 | -0.04 | 0.54 | 0.01 | -0.01 | **0.84** | |
| **UA** | 0.89 | 0.86 | 0.87 | 0.02 | 0.09 | 0.126 | -0.03 | -0.061 | 0.00 | -0.07 | -0.15 | 0.02 | **0.94** |

Table 1: Reliability, Correlation, and Discriminant Validity of Constructs

Notes: 1. AVE: Average Variance Extracted, CR: Composite Reliability, C Alpha: Cronbachs Alpha
2. ATrust: Affect-Based Trust, CTrust: Cognitive-Based Trust, INT: Purchase Intention, UA: Uncertainty Avoidance, PRV: Privacy, SEC: Security, IDV: Individualism. 3. Diagonal elements are the square root of AVE.

### Hypothesis Testing

Structural model and hypotheses were assessed by the significance of the path coefficients and the variance ($R^2$) of the dependent construct. The path significance was determined using the t-statistical calculated with the bootstrapping technique (with 500 subsamples). A five percent significance level was employed. Furthermore, the moderating effects were performed using the product indicator approach (Chin 2003). The product indicator approach refers to the product of each indicator of the independent latent construct with each indicator of the moderator construct. This approach is considered the most promising technique for the moderating effects (Henseler and Fassott 2010). Table 2 shows hypothesis testing.

|    | Path                  | Mean   | STDEV | t-statistics | p-value  |
|----|-----------------------|--------|-------|--------------|----------|
| H1 | PRV * UA -> CTrust    | 0.076  | 0.123 | 0.841        | 0.401    |
| H2 | SEC * UA -> CTrust    | 0.131  | 0.058 | 0.221        | 0.825    |
| H3 | SNS * IDV -> CTrust   | -0.110 | 0.099 | 0.467        | 0.640    |
| H4 | SNS * IDV -> ATrust   | -0.174 | 0.059 | 1.99         | 0.004*   |
| H5 | JOY * UA -> ATrust    | -0.206 | 0.131 | 2.07         | 0.039**  |
| H6 | FEAR * UA -> ATrust   | -0.169 | 0.130 | 0.723        | 0.47     |
| H7 | CTrust -> ATrust      | -0.020 | 0.060 | 1.97         | 0.049**  |
| H8 | CTrust * UA -> PINT   | -0.014 | 0.219 | 0.813        | 0.416    |
| H9 | ATrust * UA -> PINT   | 0.141  | 0.109 | 2.017        | 0.044**  |

*Table 2: path testing*

Notes: * Significant at 0.05 level, ** Significant at 0.01 level, *** Significant at 0.001 level

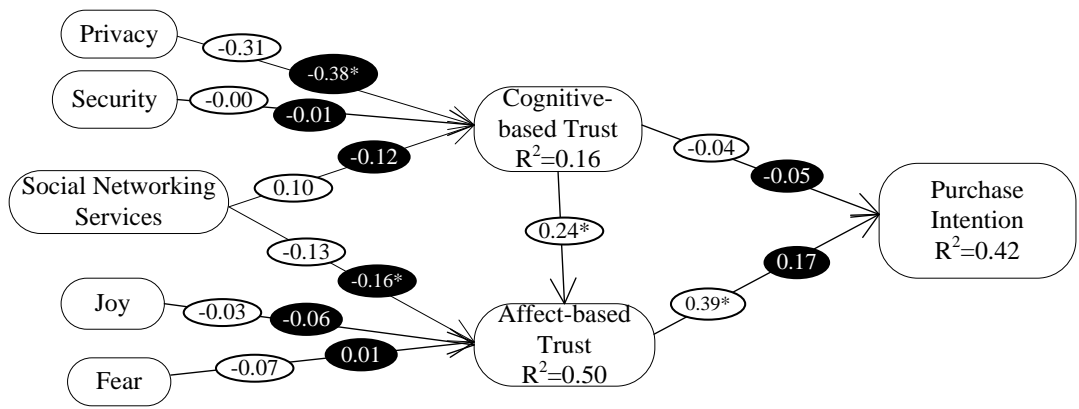

*Figure 2: path testing*　　　　　Notes: Direct Effect, Moderating effect

As shown in Table 2 and Figure 2, the results confirm the effects of moderating role of UA on (cognition, emotions), iTrust (cognitive and affect-based trust), and IDV on social networking services towards purchasing intention in B2C e-commerce at the individual level in Australia. However, the relationship is insignificant at 0.05-confidence level for H1, H2, H3, H6 and H8 while H4, H5, H7 and H9 are supported. In addition, the model shows 16 precent variance in consumer's cognitive-based trust ($R^2=0.16$) while 50 precent variance in affect-based trust ($R^2=0.50$). On the other hand, the model indicates 42 percent variance in buyer's purchase intention.

## DISCUSSION and CONCLUSION

The findings related to privacy and security concerns in a B2C website are consistent with (Kim and Kim, 2004, Kim, 2005, Gupta et al., 2010). Chen and Dibb (2010) believe that security and privacy can be guaranteed by displaying the logos of trusted third-party seals, which can assure consumers of a certain level of security protection. For instance, security assurance seals such as VeriSign, CyberTrust,

GeoTrust and Entrust guarantee consumers online transactions will be secure and have been shown to considerably affect consumers" trust in an e-commerce website (Karimov and Brengman, 2014). Interpersonal trust between consumers and online vendors could also be significantly improved if the privacy statement and security policy were displayed on every page of the website, so that online consumers are informed how they can be compensated if their financial and personal information is released to a third party.

Statistically significant support was also found for the individualism-collectivism (IDV) individual effects on the relationship between social networking services and cognitive and affect-based iTrust. This shows that consumers in Australia will consider the opinions of their peers when making their online purchase decisions. In other words, the use intensity of social networking services, such as opinions from family/friends, online consumer groups or social media cues (such as social network site, chat rooms, discussions and blogs.) is higher in building the cognitive and affect-based trust towards B2C e-commerce websites in Australia. The results suggest that social presence in e-commerce website increases consumer interpersonal trust towards purchasing online. This is in line with (Gefen and Straub, 2004, Brengman and Karimov, 2012, Hwang and Lee, 2012a).

In conclusion, the purpose of this research was to investigate how uncertainty avoidance and individualism at the individual level influences on interpersonal trust (iTrust) aspects in B2C e-commerce. Research model was proposed based on S-O-R framework; the data were collected in Australia. The findings showed that the stimulus (S) towards which a reaction is made provides a signal regarding the cognitive and affect-based trust (Organism) of an online store website, which influence consumers purchase intentions (Response). Thus, different e-business strategies would be required to establish interpersonal (cognitive and affect-based) trust between consumer and online vendors across cultures, depending on the consumer's individual cultural orientations.

### Implications

Our study extended prior research and provided essential results for research and practice. This study contributes new insights to understanding the culture values at the individual level, and how such values can significantly impact consumer's cognitive and emotional responses to interpersonal trust in B2C websites. As technology becomes an increasingly important part of e-commerce, e-vendors involved in selling products can benefit from understanding the target consumer culture, not only the country culture but also in one culturally diverse country such as Australia. The results of this study may help online shopping managers who could use the insights analysed in this research to modify their approaches. Developers and website designers can use this understanding to increase desirable outcomes by focusing the relationship between emotional reactions, cognitive evaluations and trust, to increase the chances for an online business to succeed in countries with diverse degrees of culture. Practical implications extend to business firms to make changes to their market strategies to trigger their online sale better by targeting individual consumer culture.

### Limitations and Future work

Like most survey research, this study has some limitations. First, larger sample size would have been more useful to evaluate the constancy and dependability of the findings. Second, all subjects for data collection were university students. The study could be evaluated using different group of Internet users. In addition, it will be interesting to see if the research model developed in this study is acceptable for in-group (country of origin) members living in an out-group society (country of residence). Finally, further research should determine other factors of interpersonal trust.